# The use of AR elements in the study of foreign languages at the university

Rostyslav O. Tarasenko[1][0000-0001-6258-2921], Svitlana M. Amelina[1][0000-0002-6008-3122], Yuliya M. Kazhan[2][0000-0002-4248-4248] and Olga V. Bondarenko[3][0000-0003-2356-2674]

[1] National University of Life and Environmental Sciences of Ukraine
15 Heroiv Oborony St., Kyiv, 03041, Ukraine
r_tar@nubip.edu.ua, svetlanaamelina@ukr.net
[2] Mariupol State University, 129a Budivelnykiv Ave., Mariupol, 87500 Ukraine
y.kazhan@mdu.in.ua
[3] Kryvyi Rih State Pedagogical University, 54 Gagarin Ave., Kryvyi Rih, 50086, Ukraine
bondarenko.olga@kdpu.edu.ua

**Abstract.** The article deals with the analysis of the impact of the using AR technology in the study of a foreign language by university students. It is stated out that AR technology can be a good tool for learning a foreign language. The use of elements of AR in the course of studying a foreign language, in particular in the form of virtual excursions, is proposed. Advantages of using AR technology in the study of the German language are identified, namely: the possibility of involvement of different channels of information perception, the integrity of the representation of the studied object, the faster and better memorization of new vocabulary, the development of communicative foreign language skills. The ease and accessibility of using QR codes to obtain information about the object of study from open Internet sources is shown. The results of a survey of students after virtual tours are presented. A reorientation of methodological support for the study of a foreign language at universities is proposed. Attention is drawn to the use of AR elements in order to support students with different learning styles (audio, visual, kinesthetic).

## 1 Introduction

### 1.1 The problem statement

One of the main tasks of educational institutions at the present stage is the search for new educational technologies that can help increase the efficiency of information assimilation, acquisition of professional knowledge, development of abstract thinking, the search for innovative solutions, etc. It should cause qualitative changes in the implementation of the competency-based approach to the organization of the educational process [14]. Undoubtedly, such educational technologies should be based on the use of information technologies, since their potential capabilities are inexhaustible in the processes of cognition of the surrounding world and which today can fundamentally change the traditional approaches to the presentation of learning objects, the ways of their study and research, the mapping of connections in real and virtual dimensions. One aspect of the combination of virtual and real is augmented reality (AR). Using AR technology allows a person to quickly find and receive information about real objects, which can be represented in a symbolic, sound, graphic or animated form.

In production, AR fundamentally changes the processes of designing and manufacturing technologically complex products, while increasing labor productivity and reducing errors. A special effect, as already shown by the practice of some large companies, is achieved by training personnel or improving their qualifications. In this case, first, timesaving is achieved because employees learn directly during work. In addition, the hint system is more understandable and accessible, since it can provide not only the provision of explanatory information, but even simulate the finished product based on its individual elements. Using such technologies in the professional training of specialists in higher education institutions, we can apply the latest forms of methodological support of the educational process, which will directly accompany the process of cognition and research.

The purpose of the article is to analyze the impact of the application of AR technology in the study of a foreign language by university students (using the German language as an example), to determine the advantages and possible difficulties of using this technology to develop students' foreign language communication skills.

To achieve this goal, a number of methods were used. The analysis of scientific and methodological sources showed the relevance of the issue selected for the study. Based on the comparative analysis method, the advantages of using augmented reality tools in the study of a foreign language were determined. The observation method during the execution of the task of preparing a virtual tour made it possible to see the difficulties encountered by students. The questionnaire method provided the basis for determining the attitude of students to the implementation of augmented reality elements in the study of a foreign language. The generalization method was used for a concise presentation of the research results.

### 1.2  Literature review

The technology of AR is not only increasingly used in various industries and fields of science, but attempts have already been made to apply it in the educational process. Features of using AR technology augmented in a higher education institutions are presented by Ukrainian researchers Albert A. Azaryan [24], Anna V. Iatsyshyn [7], Tetiana H. Kramarenko [10], Olena O. Lavrentieva [11], Yevhenii O. Modlo [17], Vladimir N. Morkun [15], Pavlo P. Nechypurenko [16], Vladimir N. Soloviev [18], Andrii M. Striuk [21], Elena V. Vihrova [25], Yuliia V. Yechkalo [9] and others. The use of augmented reality technology is quite common in foreign universities, and is reflected in a number of publications by scientists. In particular, according to Omer Sami Kaya and Huseyin Bicen [8], AR applications can be used in almost any educational environment, and their use in the educational process increases the level of students' knowledge.

According to Matt Bower, Cathie Howe, Nerida McCredie, Austin Robinson and David Grover [4], AR can cause a profound transformation of modern education. Overlay multimedia on the real world to see via web devices such as phones and tablet devices, means that information can be made available to students at any time and in any place. Scientists believe that this can also reduce students' cognitive overload.

Japanese researchers Marc Ericson C. Santos, Angie Chen, Takafumi Taketomi, Goshiro Yamamoto, Jun Miyazaki and Hirokazu Kato [19] identified the benefits of AR technology, which included real annotation, contextual visualization, and haptic visualization. Scientists substantiate these advantages with several latest theories – multimedia learning, experimental learning and the theory of animation visualization.

In the context of our study, the developments of scientists and practical teachers on the use of AR in the study of foreign languages are of particular interest. In particular, Robert Godwin-Jones [6] focuses on the links between AR and modern theories of foreign language learning, which emphasize localized, contextual learning and semantic connections with the real world. The researcher considers this possibility using mobile games created using the ARIS platform (AR and Interactive Storytelling), a free open source game editor of the University of Wisconsin. From his point of view, there are various ways for teachers to use the AR, because it is advisable to study the language in connection with expanded digital spaces.

Pei-Hsun Emma Liu and Ming-Kuan Tsai [12] focused on building written writing skills in English at Taiwan universities using AR through the use of multimedia documents (such as photographs and videos) in the process of learning English with computer support to improve students' language skills, which are necessary for their written assignments (writing an essay).

Murat Akçayir and Gökçe Akçayir investigated the students' attitude to their use of AR applications in learning English, in particular, for learning new vocabulary. According to the results of the study, they found that the technology saves time by simplifying the search for a new word. In addition, AR programs help students remember words. The problem that students encountered during the study, the authors indicated the recognition of the QR code. According to students, the small screens of mobile phones make it difficult to use them in teaching and learning a language [1].

Considering the search by scientists for ways to intensify the study of foreign languages and the insufficient development of this problem in terms of the use of AR technology in general and in the study of foreign languages, in addition to English, where some attempts have already been made, the problem of using AR technology in the process of learning foreign languages is relevant and requires a separate study. In addition, as the analysis of the above works shows, the application of AR technology in the study of English is mainly concentrated on the study of vocabulary, which significantly limits the use of this technology, because its potential is much greater.

## 2 Result and discussion

### 2.1 The use of elements of augmented reality for the formation of communicative competence through a virtual tour

The process of gaining knowledge usually requires the use of different methods and tools for working with information, depending on the technological possibilities and basic didactic and pedagogical models. The development of cognitive didactics has led to the emergence of a new concept of learning, based on taking into account the way people process information. At the same time, the main attention is paid to such cognitive structural and process components of learning as thinking, perception and problem solving. In the process of training aimed at obtaining new knowledge, cognitive structures should change taking into account motivational and affective factors.

New technologies, which are becoming more accessible today, contain new didactic potential regarding the possibilities of working with information in the process of studying certain topics. In particular, the study of a foreign language is impossible without the inclusion in the educational material of linguistic and geographical information related to the country of the language being studied, its traditions, the specific historical or cultural influence of the representatives of this country and the reflection of all these aspects in the students' native country or city. Since it is not always possible to carry out a real excursion to a specific region or to a particular attraction, and sometimes this is impractical due to lack of time, there is the possibility of a virtual excursion that can thematically present the contents of the excursion regardless of time, logistic and human resources. The essence of modern cognitive excursion didactics is the orientation to independent actions, which accelerates the process of acquiring knowledge. An addition, due to its specificity, the excursion has a positive motivating effect [20]. This can increase motivation to learn a foreign language, which ultimately leads to a higher efficiency of learning it.

Based on the above considerations, we chose to create a virtual tour for students learning German, concentrating on the topic "Traces of German architects in the history of Kyiv" [3]. It is worth noting that, since Kyiv is an attractive city for German-speaking tourists, several virtual tours in German have already been developed. In particular, this is the Reisen Kiew project of the Kiewer Stadtführer, which covers the most famous historical monuments of the Ukrainian capital. However, we invited students to consider the outstanding sightseeing objects of the city from a different angle, namely, as indicated in the topic – in terms of the contribution of German architects to their design and construction.

At the initial stage, the selection of objects for a virtual tour was carried out. For this purpose, a number of materials were analyzed regarding historical objects in the territory of the city of Kyiv. The following architectural monuments were selected:

1. St. Volodymyr's Cathedral. The construction of the cathedral began in 1862 and lasted 40 years. Its construction involved several architects and painters. In 1853-1859, the prominent architect of German origin, Ivan Strom, designed the St. Volodymyr's Cathedral; architects P. Sparro, A. Beretti and V. Nikolaev amended the design. Later, German engineer Berengardt was involved in solving technical problems.
2. St. Sophia's Cathedral. The cathedral, built in 1037, was destroyed several times. In 1736-1740, the Ukrainian architect of North German origin, Johann Gottfried Schedel reconstructed the main bell tower. He also built a stone wall around the St. Sophia's Monastery, very successfully combining Western style elements with elements of the Cossack Baroque and folk motifs.
3. Kyiv Pechersk Lavra. Until 1745, the architect and engineer Johann Gottfried Schedel worked on the construction of the bell tower of the Kyiv Pechersk Lavra, which became one of the best bell towers in Eastern Europe of the 18th century. Schedel developed a project in a transitional style from baroque to classicism. The bell tower of the Assumption Cathedral was built according to his design in the form of an octagonal four-tier tower with a height of 96.5 meters.
4. St. Andrew's Church. The foundations of St. Andrew's Church were built according to the design of I.G. Schedel; however, the design of the temple itself, submitted by him, was not approved. Carved details of the iconostasis, according to sketches and drawings by F.-B. Rastrelli, created by the master (J. Domash, A. Karlovsky, M. Manturov, D. Ustars, H. Oreidah, J. Zunfer), among which there were several Germans. German master Johann Friedrich Grot led installation work.
5. National Opera of Ukraine. After the old theater building burned down in 1896, an international architectural competition for the design of a new opera house was announced. More than twenty well-known architects from different countries – Italy, Germany, Russia and France – attended the competition, and the winner was the project of the architect of German-Baltic origin Victor Schröter, a representative of the rational direction of eclecticism in architecture. The new city theater was built from 1898 to 1901 in the style of rationalism, baroque and neo-Romanesque style.
6. Klov Palace. The architects J. G. Schedel and P. I. Neyelov built Klov Palace in 1756. The German painter and jeweler Benedict Friedrich performed a number of works, in particular, the painting of the ceiling in the large hall of the Klov Palace. The German garden master Johann Blech worked on the Klovsky garden.
7. Kyiv Polytechnic Institute. Famous architects took part in the competition for construction projects at the Polytechnic Institute, including Germans and Austrians, in particular: Benoit, Gauguin, Kitner, Kobelev, Pomerantsev, Tsender and Schröter. The jury recognized the best project of Professor I. S. Kitner, under the motto "Prestissimo" ("Very Fast"). The construction of six university buildings in the Romanesque style began on August 30, 1898 and was completed in 1901.

After determining the content of the future virtual tour, that is, the selection of the outstanding architectural structures of Kyiv associated with the work of German architects, engineers and decoration painters, information resources were identified that students can use to prepare and conduct a virtual tour. Providing students with assistance in information resources was determined, on the one hand, by the desire to reduce the time for them to complete the task, since local history aspects are only part of the German language classes, and, on the other hand, to limit the amount of information for processing by directing it to certain subtopics. Interactivity, a variety of materials and multimedia play an important role in creating a virtual tour. Another important aspect was also the understanding that when integrating information into a virtual tour, we should respect copyrights, that is, use only those sources that are publicly available or those for which a permit is granted.

First, we suggested that students include the use of a digital map of Kiev in the structure of a virtual tour, since the maps provide an understanding of the integrity of the territory with objects located on it and possible connections between them. They form a sense of scale and improve spatial orientation. Using digital maps, students can easily create virtual sightseeing tours, combining sightseeing objects with routes according to certain signs: the chosen topic, the chronological period, the place of a historical event, the sequence of location, the logic of movement. In our study, we used the Google Maps application as a tool for creating a virtual tour map. One of the advantages of this tool is the ability to clear the position of the excursion object on the map using built-in search tools based on addresses. Colored markers were superimposed on automatically identified points on a digital map to conveniently identify each virtual tour object. At the location of the excursion objects, we applied color marking for convenient use (fig. 1).

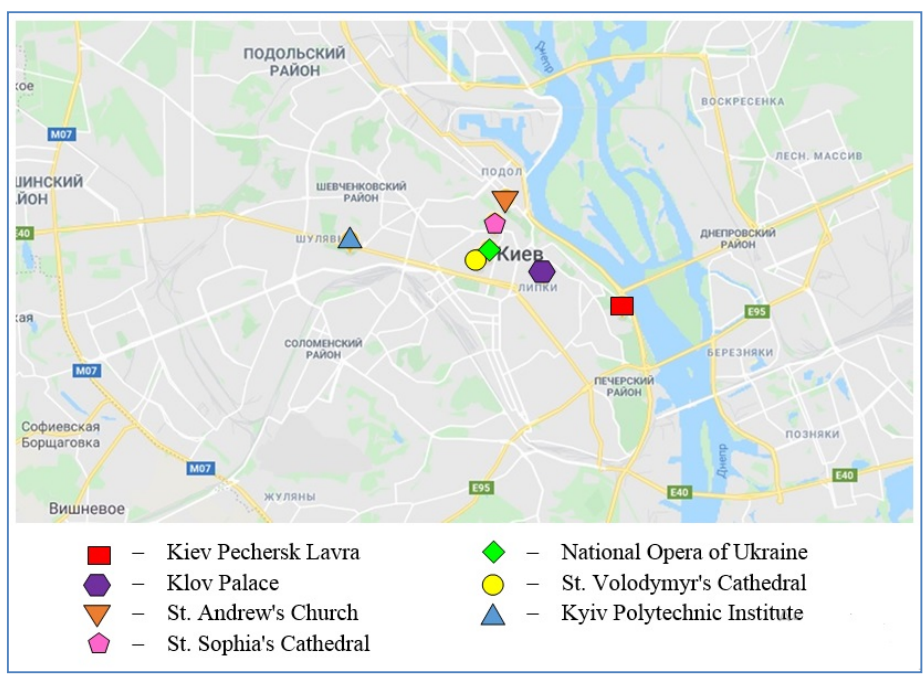

**Fig. 1.** Digital map of Kyiv with printed objects proposed for a virtual tour (Google resource).

The main task of students was to develop their own excursions based on the use of the proposed map. At the same time, each group selects one of the characteristics for building the route. As already noted, the virtual tour was to maximize the achievement of the main goal, in particular, the deepening of the study of the German language by acquaintance with architectural monuments built with the participation of German architects. In this case, the informative part about the objects of the virtual excursion had to combine text, photo and video information into a single, complementary information case [2; 22; 23], formed using AR technologies. Guided by these conditions, access to the necessary information on mapped architectural monuments should be provided throughout the tour. One of the ways to obtain information, quickly and conveniently, in various forms is the use of modern mobile devices that are capable of reproducing multimedia information concentrated on various web pages. An important issue remains the search for the right information and quick access.

We asked students to solve this problem by creating a system of QR codes that provide information support for a virtual tour, providing quick access to information about a particular object of the tour in different forms. It is known that a QR code can be generated for textual information, a URL, an e-mail, a phone number, etc. They can be easily and stably scanned by special scanners and provides quick access to encoded information.

At the initial stage, we conducted a training for students to develop the skills of generating QR codes for these types of information using freely available systems. After that, students processed open Internet resources with text, photo and video information about the objects of the excursion, selected the most successful of them, and then, using QR-code generators, formed the corresponding set of codes. We show an example of a set of QR codes for information about one of such excursion objects, the bell tower of the Kyiv Pechersk Lavra (fig. 2).

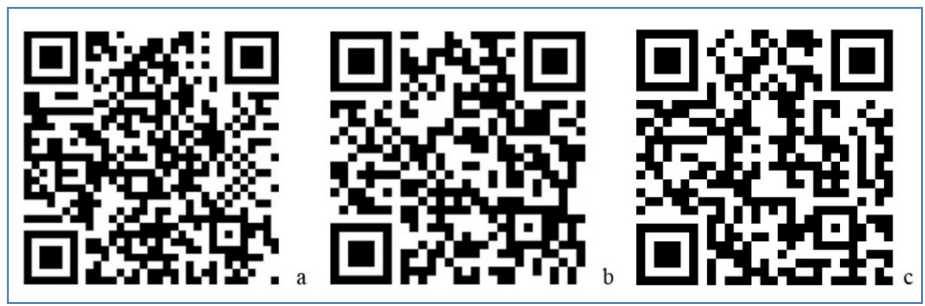

**Fig. 2.** A set of QR codes with text (a), graphic (b) and video (c) about the bell tower of the Kyiv Pechersk Lavra.

The main condition for the preparation of textual information was that it should be in German. One of the sources that students used for this purpose was the open electronic encyclopedia Wikipedia (fig. 3). This approach had a double effect, since students, on the one hand, processed German sources in the process of searching and selecting the necessary information, and on the other hand, created the opportunity to receive extended information in German about objects during the virtual tour for her "visitors", which were students from other groups.

However, for many people, information in the form of a graphic image is more informative than text. In particular, many facts can be presented more fully and clearly in the photograph than in words. Therefore, in a virtual tour the use of images is especially important. In order for the image to be used in a virtual tour, they must be presented in digital form. The range of such images can be very diverse and range from simple photographs to interactive maps, managed panoramic images, 3D images and the like. Image types such as satellite images are also well suited for inclusion in virtual tours. The use of mobile devices in the process of conducting virtual excursions with access to images about the object has significant advantages compared to providing these images in print, primarily due to the possibility of increasing images, changing their brightness and contrast, making even small details visible. When preparing virtual excursions, students sought to provide access through a QR code not to individual images about the object, but to a collection of photographs that would allow them to get the most out of a particular architectural landmark (fig. 4). For this purpose, students used the resources of Google Images, Wikiway and the like.

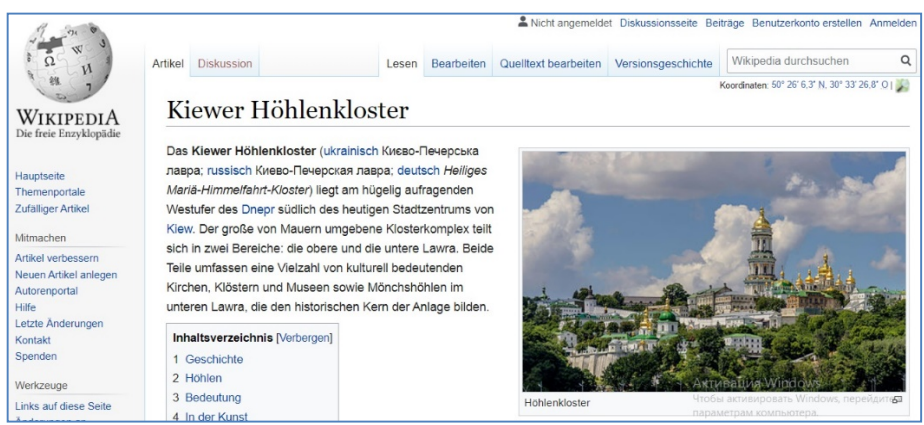

**Fig. 3.** A fragment of a web page with textual information about the bell tower of the Kyiv Pechersk Lavra, access is generated by a QR code.

The advantages of video resources are that the presentation of information on the corresponding excursion space is almost realistic and relatively uncomplicated. Like photographs, especially panoramic photographs, films and videos very closely convey the atmosphere of real visits to places of excursion objects. In addition, in the case of using video, there is not only visual perception, but also perception of information by ear.

On this basis, when designing virtual excursions, students integrated thematic films and videos by linking to video portals such as You Tube and Google Video. We show an example of a link to the corresponding video fragment encoded by a QR code (fig. 5). Topically relevant videos can be quickly found using targeted keyword searches. As with photographs, we also need to respect copyrights regarding videos and films.

It is worth noting that the communication of the group members during the preparation of the virtual tour also contributed to the formation of teamwork skills among students and provided them with the opportunity to develop communication skills in foreign languages. In addition, working together on one topic and intending to achieve a common goal, students learned from each other to build sentences of different types with the correct word order, learned conversational vocabulary options, trained pronunciation of individual words and phrases.

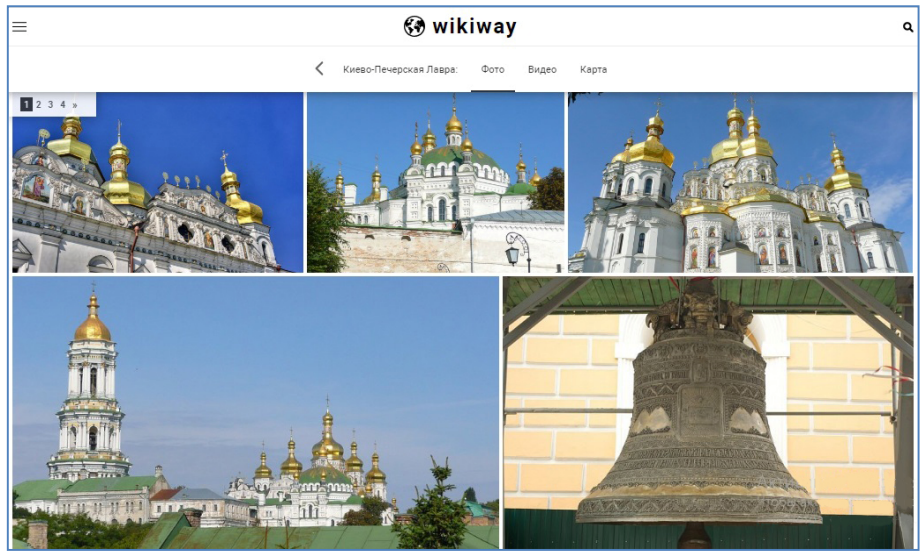

**Fig. 4.** A fragment of a web page with graphic information about the bell tower of the Kyiv Pechersk Lavra, access is generated by a QR code.

### 2.2 Augmented reality as a modern educational solution for studying foreign languages

Upon completion of the development of virtual tours of each group, students who did not take part in their preparation tested them. After passing these excursions, a questionnaire was proposed, which was aimed at assessing the effectiveness of a virtual excursion with elements of AR in studying the German language. This questionnaire contained questions grouped into four blocks: motivational, informative, linguistic and technological. 39 people attended the survey. The results of the answers to the questionnaire are shown in table 1.

The results of the survey indicate that the use of virtual excursions with elements of AR aroused interest among students, which manifested itself to different degrees and in different aspects when studying the German language. In particular, this approach has most positively affected the substantive aspect of this process. A rather high percentage of students (76.9%) noted that the elements of AR provided them with extended information about the excursion objects presented.

It is gratifying to note that the level of positive answers in the technological unit was also quite high (58.1%), which indicates students' readiness for new forms of organizing the study of a foreign language. However, some aspects of this process caused quite serious technological difficulties. In particular, 61.5% of students were not able to use fully the capabilities of the proposed elements of AR due to insufficient technical characteristics and an inappropriate software set for their own smartphones.

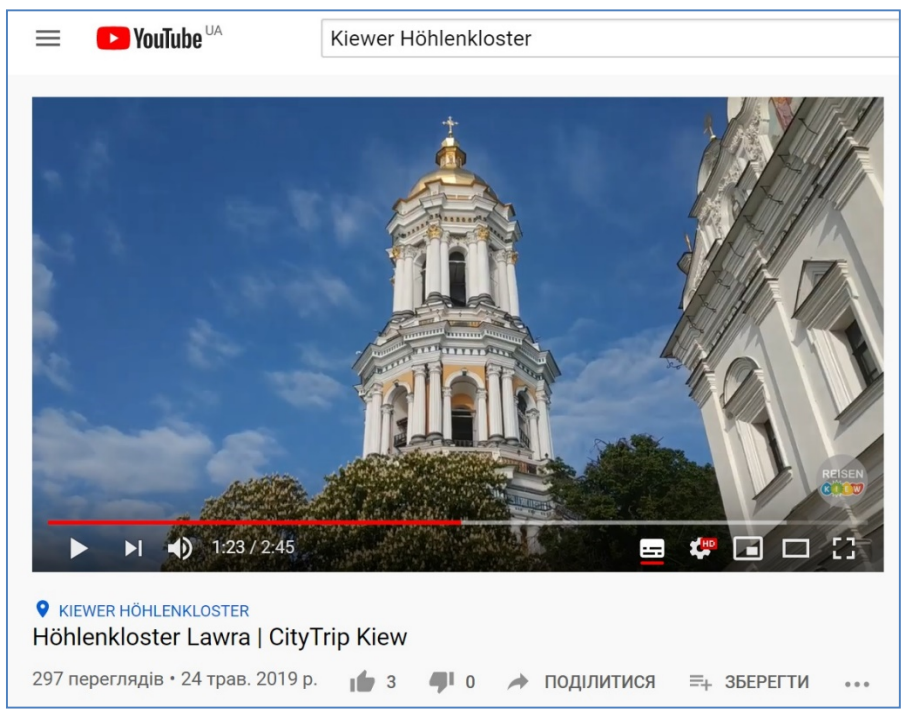

**Fig. 5.** A fragment of a web page with video information about the bell tower of the Kyiv Pechersk Lavra, access is generated by a QR code (Reisen Kiew).

An undoubtedly positive result of using virtual tours is the desire expressed by 79.5% of students to learn German, including in this way. Therefore, it is advisable for teachers to use the influence on the motivation to learn a foreign language, which is created using AR elements in the educational process.

Another confirmation of the advisability of using elements of AR in the study of a foreign language is the low level of positive answers to the questions of the linguistic block of the questionnaire. This indicates that the general level of students' linguistic knowledge is quite low and therefore needs to be improved, including through the search for new approaches and forms of learning a foreign language.

Thus, the use of AR technology contains great potential for the formation of a holistic, realistic view of objects outside the classroom. Owing to the student's independent actions and his emotional impression, when perceiving the educational object, an active approach of the educational content to the student occurs, which leads to better assimilation and longer memorization of knowledge.

Improving the effectiveness of training and longer memorization of the studied content is achieved through higher motivation for learning, active, and direct interaction with a real educational object based on AR technology. Since there are different types of students depending on the channel of perception of information (audial, visual, kinesthetic, mixed types, etc.) [5; 13], thanks to the holistic representation of objects based on AR technology, a higher level of assimilation of educational information and the formation of multimodal representations can be achieved.

**Table 1.** Results of answers to questionnaire questions.

| Question | Response rate | |
|---|---|---|
| | **Yes** | **No** |
| **Motivational block** | **66.7** | **33.3** |
| Did the virtual tour contribute to the desire to learn German? | 79.5 | 20.5 |
| Are you ready to continue learning the language this way? | 66.7 | 33.3 |
| Have you been encouraged by the existing elements of AR to depth study of information in German about the excursion objects presented? | 53.8 | 46.2 |
| **Content block** | **77.8** | **22.2** |
| Have elements of AR provided you with enhanced information about the excursion objects presented? | 76.9 | 23.1 |
| Did German videos provide understanding of the information about the object of the excursion? | 74.4 | 25.6 |
| Were the text materials sufficient to obtain information on the topic of the tour? | 82.1 | 17.9 |
| **Linguistic block** | **53.8** | **46.2** |
| Did the information presented in the form of elements of AR make it easier for you to understand excursion materials in German? | 61.5 | 38.5 |
| Did elements of AR help to understand the meaning of new words in context? | 56.4 | 43.6 |
| Have elements of augmented reality contributed to better memorization of terms? | 43.6 | 56.4 |
| **Technological block** | **58.1** | **41.9** |
| Were there new ways for you to obtain additional information using QR codes? | 71.8 | 28.2 |
| Have you possessed sufficient skills in using smartphones to receive information presented as elements of AR? | 64.1 | 35.9 |
| Did the specifications and software set of smartphones make it possible to utilize fully the capabilities of the proposed elements of AR? | 38.5 | 61.5 |

The use of AR technology requires appropriate methodological didactic reorientation, which will create the opportunity for students to independently organize research, collect, evaluate, process and present information, apply complex hypertext structures, develop network thinking, work within flexible, group, project-oriented forms of training.

## 3      Conclusions

In the course of the study, a number of advantages of using AR technology in the study of the German language were identified. In our opinion, such advantages can be used in the process of learning other foreign languages, in particular:

- The technology of AR allows to achieve a higher level of assimilation of educational material, since various channels of perceiving information are involved, because it is important for studying a foreign language what type of perception of information the student belongs to, whether he is an audial, visual, kinesthetic, and the like.
- Due to the integrity of the representation of the studied object, the student can get a more complete picture of it, and then learn, for example, a larger amount of new lexical material, since memorizing new words, especially terminology, takes place faster and remains in memory longer when new words are not used in isolation, but in context.
- Based on the application of AR technology, students can familiarize themselves with objects, unique or inaccessible due to spatial remoteness (for example, located in another country). It helps them to understand the essence or purpose of these objects and to remember the vocabulary associated with them, which would be much more difficult to use other information sources.
- Faster memorization of new vocabulary is also facilitated by the parallel presentation of information case together with selected objects for study, which allows students to quickly receive extended information using AR technologies.
- The use of AR technology, in particular in the form of a virtual tour, which involves working in a group, allows students to develop communicative foreign language skills.
- AR technology can be a good tool for learning a foreign language, because it allows the student to learn at his own pace. The assimilation of new knowledge and skills takes place based on previous knowledge of the language, the level of which, as shown by pedagogical practice, is very different even within the same academic group.

The main problem of using AR technology when learning a foreign language by organizing virtual tours, in our study, as in other cases of using digital information, is the dependence on the technical infrastructure and software. Since each student used his own smartphone with different technical characteristics and his own software set, sometimes this led to problems with receiving and reproducing information in accordance with the used technology. Most of these problems were related to ensuring stable access to the Internet, improper operation of QR scanners and the lack of some software installation skills.

In order to understand better the transfer of knowledge through virtual and AR and to be able to develop appropriate methods for using these technologies, further research is needed. In particular, it is advisable to compare augmented and virtual reality technologies with traditional teaching methods and other latest information processing tools, as well as study and compare various methods that offer augmented and virtual reality.

142